\documentclass[final,5p,times,twocolumn]{elsarticle}

\usepackage{lineno}
\usepackage[hidelinks]{hyperref}

\usepackage{graphicx}
\usepackage{subfigure}
\usepackage{amsmath}   

\begin{document}

\begin{frontmatter}

\title{Preliminary design of a Cavity Tuner for Superconducting Radio-Frequency Cavity}

\author[1,2]{Ming Liu}

\author[1,2]{Jiyuan Zhai}

\author[1,2]{Feisi He}

\author[1,2]{Zhenghui Mi\corref{cor1}%
        }\ead{mizh@ihep.ac.cn}

 \cortext[cor1]{Corresponding author}

\address[1]{Institute of High Energy Physics, Chinese Academy of Sciences, Beijing 100049, China}
\address[2]{University of Chinese Academy of Sciences, Beijing 100049, China}

\begin{abstract}
This paper introduces a newly designed cavity tuner for superconducting radio-frequency (SRF) cavity. Aiming to overcome the drawbacks of traditional tuning systems, like the limited tuning range of piezoelectric tuner and the low-speed tuning of stepper-motor-based tuner, this novel tuner is crafted to improve SRF cavity performance and stability via efficient and accurate frequency tuning. The design encompasses several key elements. The cavity structure includes a commonly used 1.3 GHz single-cell superconducting cavity and a room-temperature coaxial tuner cavity. The coupling mechanism between the two cavities, along with the coupling window design, ensures effective energy transfer while minimizing losses. The mechanical tuning system, driven by electromagnetic coils, enables precise adjustments, and the cooling mechanisms for both cavities guarantee stable operation. Functioning by coupling an external resonant cavity to the superconducting one, this tuner can adjust frequencies through mechanical or electromagnetic methods. It realizes rapid tuning, with a speed much faster than traditional mechanical tuner, high-precision tuning down to the sub-mHz level, and a wide tuning range covering a broader frequency spectrum. Theoretical analysis and simulations verify that the tuner can remarkably enhance tuning speed, precision, and range. It also has distinct advantages such as a simplified structure, which reduces manufacturing and maintenance complexity, and enhanced reliability due to its non-contact tuning operation. In particle accelerators, this cavity tuner holds great potential. It represents a significant step forward in superconducting accelerator technology, offering a novel way to optimize the performance and stability of SRF cavity.
\end{abstract}


\end{frontmatter}


\section{Introduction}
\label{sec1}

High-energy particle accelerators serve as indispensable scientific research tools for probing the properties of fundamental particles and uncovering the mysteries of the universe. At the heart of these devices are superconducting accelerators, which rely on superconducting cavities. Superconducting cavities are the linchpin components that determine the performance of the accelerator. Their performance is gauged by crucial parameters such as the acceleration gradient and the quality factor (Q - value)\cite{padamsee_superconducting_2023}. 

The acceleration gradient, for instance, reflects the cavity's ability to accelerate particles. A higher acceleration gradient can substantially reduce the number of required cavities. In some large - scale accelerator projects, increasing the gradient from 10 MV/m to 20 MV/m could potentially halve the number of cavities, thereby significantly lowering the accelerator's construction cost\cite{aghayan_development_2021}. The quality factor, on the other hand, assesses the cavity's energy loss. A cavity with a Q - value of \(10^9\) can consume up to 50\% less energy at least compared to one with a Q - value of \(10^8\), leading to substantial savings in operational costs\cite{Kolb2013}.

The performance of superconducting cavity is subject to multiple influencing factors, including geometric shape, material properties, and post-processing methods. Moreover, the operational mode of the cavity plays a pivotal role. Superconducting cavity can operate in either continuous or pulsed modes. In continuous mode, a higher external quality factor (\(Q_e\)) is set to enhance the energy transfer efficiency to the beam. However, this mode demands more power from the source. For example, in a continuous mode accelerator, the power source may need to supply 500 kW compared to 200 kW in pulsed mode\cite{Steder:619223}. Additionally, it increases the dynamic heat load on the cryogenic system, which can raise cooling costs. Pulsed mode reduces the power source and cryogenic system costs. Nevertheless, it is limited by dynamic Lorentz force detuning, which can cause ponderomotive effects and restrict the \(Q_e\)\cite{QIU2021165633}.

To enhance the performance and stability of superconducting cavity, a tuning system is essential. Traditional tuning systems are generally classified into fast tuners and slow tuners. Fast tuners, which utilize piezoelectric ceramics, can achieve rapid and precise tuning. However, their tuning range is typically limited to within ±50 kHz. Slow tuners, driven by stepper motors, offer a broader tuning range, often up to ±1 MHz, but they are slower and less precise, with a response time of several seconds compared to milliseconds for fast tuners\cite{contrerasmartinez2022testsextendedrangesrf}. 

To integrate the advantages of both, at the SRF2021 conference, Kuzikov from Fermilab proposed a novel cavity tuner\cite{Kuzikov2022}. This tuner couples an external resonant cavity to the superconducting cavity, enabling frequency adjustment by tuning the external cavity. This innovative design can achieve rapid, precise, and wide-range tuning, which is expected to significantly enhance the performance and stability of superconducting cavity.

Building on the 1.3 GHz single-cell superconducting cavity design widely used in projects like the International Linear Collider (ILC) and X-ray Free-Electron Laser (XFEL), this work focuses on developing a cavity tuner tailored for this representative SRF cavity type. The 1.3 GHz single-cell cavity, characterized by its balance of accelerating gradient and compact structure, has been proven effective in high-energy electron accelerators but faces challenges in dynamic frequency stability under varying operational modes. By integrating an external coaxial tuner cavity (room-temperature, vacuum-compatible) with the 1.3 GHz superconducting cavity, this study aims to address the limitations of conventional tuners while leveraging the cavity’s intrinsic high Q-value potential (e.g.,  Q$_{0}$ $>$ 10$^{10}$ after medium-temperature baking\cite{He_2021,YANG20221354092,PhysRevAccelBeams.27.092003}).

\section{Principle of cavity tuner}
\label{sec2}

The electromagnetic field distributions in a resonant cavity can be expressed as\cite{slater1950microwave}:
\begin{equation}
    \vec{E}(\vec{r}, t) = \operatorname{Re}\left\{\sum_{m} \hat{\mathcal{E}}^{(m)} \vec{e}^{(m)}(\vec{r}) e^{i \omega^{(m)} t}\right\}
    \label{eq:elecfield}
\end{equation}
\begin{equation}
    \vec{H}(\vec{r}, t) = \operatorname{Re}\left\{\sum_{m} \hat{\mathcal{H}}^{(m)} \vec{h}^{(m)}(\vec{r}) e^{i \omega^{(m)} t + i \pi/2}\right\}
    \label{eq:magfield}
\end{equation}
where $\hat{\mathcal{E}}^{(m)}$ and $\hat{\mathcal{H}}^{(m)}$ denote the complex amplitudes, $\vec{e}^{(m)}(\vec{r})$ and $\vec{h}^{(m)}(\vec{r})$ represent the spatial distributions of the $m$-th electromagnetic mode, and $\omega^{(m)}$ is the corresponding resonant frequency.

The time-domain relationship between electric and magnetic fields in the resonant cavity is described by\cite{Liepe2001ts}:
\begin{equation}
    \begin{split}
        \frac{d^2}{dt^2}\mathcal{E}^{(m)}(t) + \left(\omega^{(m)}\right)^2\mathcal{E}^{(m)}(t) 
        &= -\frac{1}{\varepsilon_0}\frac{d}{dt}\int_V \vec{j}\cdot\vec{e}^{(m)} \, dv \\
        &\quad - c\omega^{(m)}\int_S (\vec{n}\times\vec{E}_{\text{tan}})\cdot\vec{h}^{(m)} \, ds
    \end{split}
    \label{eq:elecfieldtime}
\end{equation}
\begin{equation}
    \begin{split}
        \frac{d^2}{dt^2}\mathcal{H}^{(m)}(t) + \left(\omega^{(m)}\right)^2\mathcal{H}^{(m)}(t) 
        &= c\omega^{(m)}\int_V \vec{j}\cdot\vec{e}^{(m)} \, dv \\
        &\quad - \frac{1}{\mu_0}\frac{d}{dt}\int_S (\vec{n}\times\vec{E}_{\text{tan}})\cdot\vec{h}^{(m)} \, ds
    \end{split}
    \label{eq:magfieldtime}
\end{equation}
where $\vec{E}_{\text{tan}}$ is the tangential electric field on the cavity surface, $\vec{n}$ is the unit normal vector of the cavity wall, and $c$ is the speed of light in vacuum.

\begin{figure}[htbp]
    \centering
    \includegraphics[width=0.45\textwidth]{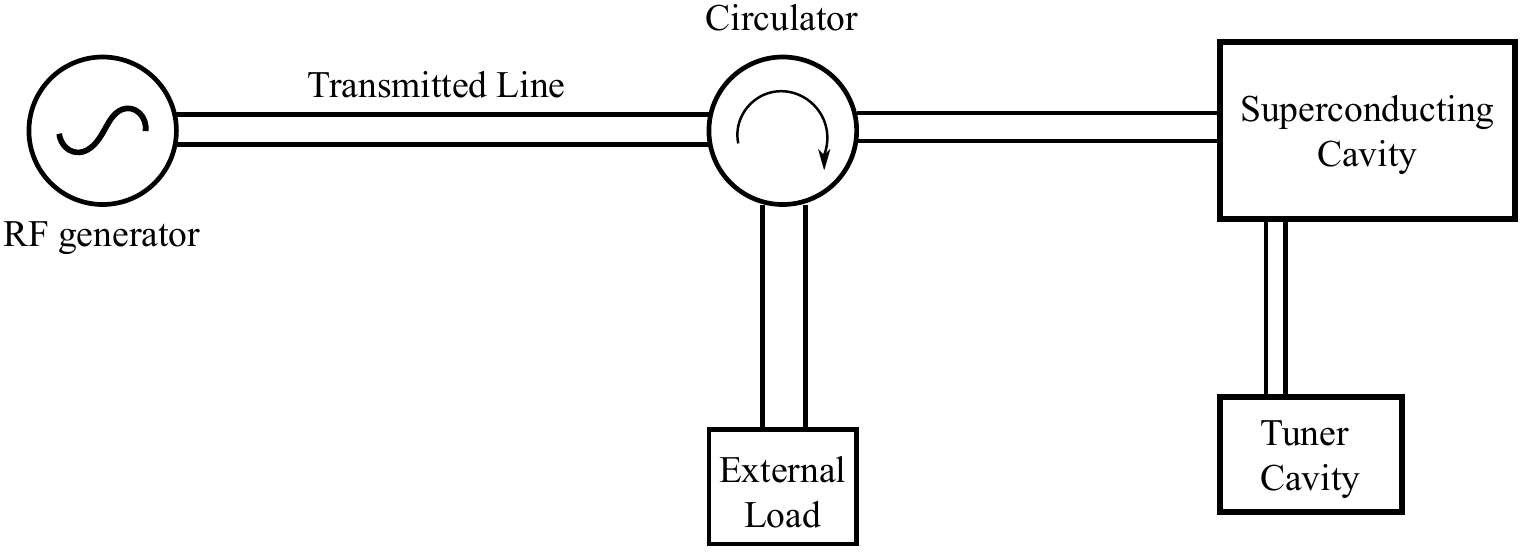}
    \caption{Schematic diagram of the cavity tuner system, illustrating the coupling between superconducting cavity and tuning cavity through a transmission line}
    \label{fig:schematic}
\end{figure}

For the cavity tuner system (schematic shown in Fig. \ref{fig:schematic}), the RF power source delivers energy to the superconducting cavity through a transmission line and circulator. The circulator routes reflected power to an external load, protecting the power source from damage. The superconducting cavity and tuning cavity are coupled via a short transmission line, enabling energy exchange between the two. The resonant frequency of the tuning cavity can be adjusted by two methods: 
1. Mechanical tuning: altering the cavity length using a bellows-based mechanism (see Section \ref{sec:mechanicaltuning}); 
2. Dielectric tuning: changing the effective permittivity of the intracavity medium\cite{PhysRevAccelBeams.27.052001}.

The current density excited by the RF power source at the coupling port is expressed as:
\begin{equation}
    \vec{j}_g(t,\vec{r}) = \operatorname{Re}\left\{ J_g(t) \vec{f}_g(\vec{r}) \right\}
    \label{eq:current}
\end{equation}
where $J_g = \hat{J}_g e^{i \omega_g t}$ represents the time-varying complex amplitude, and $\vec{f}_g(\vec{r})$ is the spatial distribution function of the excitation source. Substituting this into the electric field equation \eqref{eq:elecfieldtime}, the power source term becomes:
\begin{equation}
    \frac{1}{\varepsilon_0}\frac{d}{dt}\int_V \vec{j}_g \cdot \vec{e}^{(m)} dv = \frac{1}{\varepsilon_0}\frac{d}{dt}J_g(t) \int_V \vec{f}_g(\vec{r}) \cdot \vec{e}^{(m)}(\vec{r}) \, dv
    \label{eq:currenttime}
\end{equation}
where the spatial overlap integral $\int_V \vec{f}_g(\vec{r}) \cdot \vec{e}^{(m)}(\vec{r}) \, dv$ characterizes the modal coupling efficiency between the current source distribution and the $m$-th eigenmode of the resonant cavity. This integral essentially represents the projection of the source current distribution onto the modal field pattern.

The coupling coefficient of the coupler is defined as:
\begin{equation}
    K_{g}^{(m)} = \int_{V} \vec{f}_{g}(\vec{r}) \cdot \vec{e}^{(m)}(\vec{r}) \, \mathrm{d}v
    \label{eq:couplercouplingfactor}
\end{equation}
This coefficient \(K_{g}^{(m)}\) quantifies the coupling strength between the power source and the \(m\)-th mode of the cavity. A larger value of \(K_{g}^{(m)}\) implies a more efficient energy transfer from the power source to the corresponding mode within the cavity.

This coefficient \(K_{g}^{(m)}\) determines the energy transfer efficiency from the external power source to the $m$-th cavity mode through spatial mode matching. The magnitude of \(K_{g}^{(m)}\) directly correlates with the excitation effectiveness: values approaching unity indicate optimal impedance matching and maximal energy transfer, whereas smaller values correspond to weaker modal excitation.

Moreover, the current density due to external losses can be expressed as:
\begin{equation}
    \vec{j}_{c}(t, \vec{r}) = \sigma_{c}(\vec{r}) \vec{E}(t, \vec{r}) = \sigma_{c}(\vec{r}) \sum_{k} \mathcal{E}^{(k)}(t) \vec{e}^{(k)}(\vec{r})
    \label{eq:externalcurrent}
\end{equation}
where \(\sigma_{c}(\vec{r})\) represents the conductivity distribution related to external losses at position \(\vec{r}\) within the cavity. This equation shows that the external - loss current density is linearly proportional to the electric field and the local conductivity.

Consequently, the external loss term corresponding to the first term on the right - hand side of Equation \eqref{eq:elecfieldtime} can be derived as follows:
\begin{equation}
    \frac{1}{\varepsilon_{0}}\frac{\mathrm{d}}{\mathrm{d}t}\int_{V} \vec{j}_{c} \cdot \vec{e}^{(m)} \, \mathrm{d}v = \frac{1}{\varepsilon_{0}} \sum_{k} \frac{\mathrm{d}\mathcal{E}^{(k)}(t)}{\mathrm{d}t} \int_{V} \sigma_{c}(\vec{r}) \vec{e}^{(k)}(\vec{r}) \cdot \vec{e}^{(m)}(\vec{r}) \, \mathrm{d}v
    \label{eq:externalloss}
\end{equation}
This equation describes how the time - varying external - loss current contributes to the overall change in the \(m\) - th mode of the electric field within the cavity.

The external quality factor is defined as:
\begin{equation}
    \frac{1}{Q_{e}^{(km)}} := \frac{1}{\varepsilon_{0} \omega^{(m)}} \int_{V} \sigma_{c}(\vec{r}) \vec{e}^{(k)}(\vec{r}) \cdot \vec{e}^{(m)}(\vec{r}) \, \mathrm{d}v
    \label{eq:externalQ}
\end{equation}
The external quality factor \(Q_{e}^{(km)}\) reflects the ratio of the energy stored in the cavity to the energy dissipated due to external losses for the interaction between the \(k\) - th and \(m\) - th modes. A higher \(Q_{e}^{(km)}\) value indicates lower external losses and more efficient energy storage within the cavity.

Due to the influence of cavity wall loss \cite{bafia2025signaturesenhancedsuperconductingproperties}, the second term on the right - hand side of Equation \ref{eq:elecfieldtime} can be expressed as:
\begin{equation}
    \begin{split}
        \int_{S} (\vec{n} \times \vec{E}_{\text{tan}}) \cdot \vec{h}^{(m)} \, \mathrm{d}s 
        &= (1 + i)R_{s}^{(m)}\mathcal{H}^{(m)}(t) \int_{S} \vec{h}^{(m)} \cdot \vec{h}^{(m)} \, \mathrm{d}s \\
        &= (1 + i)\mu_{0} \frac{\omega^{(m)}}{Q_{0}^{(m)}}\mathcal{H}^{(m)}(t)
    \end{split}
    \label{eq:wallloss}
\end{equation}
where \(R_{s}^{(m)} := \sqrt{\frac{\mu_{0} \omega^{(m)}}{2 \sigma}} = \frac{1}{\sigma \delta_{s}} \propto \frac{(\omega^{(m)})^2}{T} e^{-1.76 T_{c} / T}\) represents the surface resistance of the superconducting cavity. Here, \(\mu_{0}\) is the vacuum permeability, \(\sigma\) is the conductivity, \(\delta_{s}\) is the skin depth, \(T\) is the temperature, and \(T_{c}\) is the critical temperature of the superconductor.

In summary, Equation \ref{eq:elecfieldtime} can be rewritten as:
\begin{equation}
    \begin{aligned}
        \frac{\mathrm{d}^2 \mathcal{E}^{(m)}}{\mathrm{d}t^2} 
        &+ \frac{\omega^{(m)}}{Q_{0}^{(m)}}(1 + i)\frac{\mathrm{d} \mathcal{E}^{(m)}}{\mathrm{d}t} 
        + \sum_{k} \frac{\omega^{(m)}}{Q_{e}^{(km)}} \left\{ \frac{\mathrm{d} \mathcal{E}^{(k)}}{\mathrm{d}t} 
        + \frac{\omega^{(m)}}{Q_{0}^{(m)}}(1 + i)\mathcal{E}^{(k)} \right\} \\
        &+ (\omega^{(m)})^2 \mathcal{E}^{(m)} 
        = -\frac{K_{g}^{(m)}}{\varepsilon_{0}} \left\{ \frac{\mathrm{d}}{\mathrm{d}t}J_{g} 
        + \frac{\omega^{(m)}}{Q_{0}^{(m)}}(1 + i)J_{g} \right\} 
    \end{aligned}
    \label{eq:final}
\end{equation}

Given that the RF cavity satisfies \(1/Q_{e}^{(km)} \ll 1\) and \(1/Q_{0}^{(m)} \ll 1\), the terms \(\sum_{k} \frac{\omega^{(m)}}{Q_{e}^{(km)}} \cdot \frac{\omega^{(m)}}{Q_{0}^{(m)}} (1 + i) \mathcal{E}^{(k)}\) and \(-\frac{K_{g}}{\varepsilon_{0}} \cdot \frac{\omega^{(m)}}{Q_{0}^{(m)}} (1 + i) J_{g}\) in the above equation have sufficiently small coefficients. Therefore, these terms can be regarded as higher - order infinitesimals, and the equation can be simplified as follows:
\begin{equation}
    \begin{aligned}
        \frac{\mathrm{d}^2 \mathcal{E}^{(m)}}{\mathrm{d}t^2} 
        &+ \frac{\omega^{(m)}}{Q_{0}^{(m)}}(1 + i)\frac{\mathrm{d} \mathcal{E}^{(m)}}{\mathrm{d}t} 
        + \sum_{k} \frac{\omega^{(m)}}{Q_{e}^{(km)}}\frac{\mathrm{d} \mathcal{E}^{(k)}}{\mathrm{d}t} \\
        &+ (\omega^{(m)})^2 \mathcal{E}^{(m)} 
        = -\frac{K_{g}^{(m)}}{\varepsilon_{0}} \frac{\mathrm{d}}{\mathrm{d}t}J_{g}
    \end{aligned}
    \label{eq:simplified}
\end{equation}

\begin{figure}[htbp]
    \centering
    \includegraphics[width=0.45\textwidth]{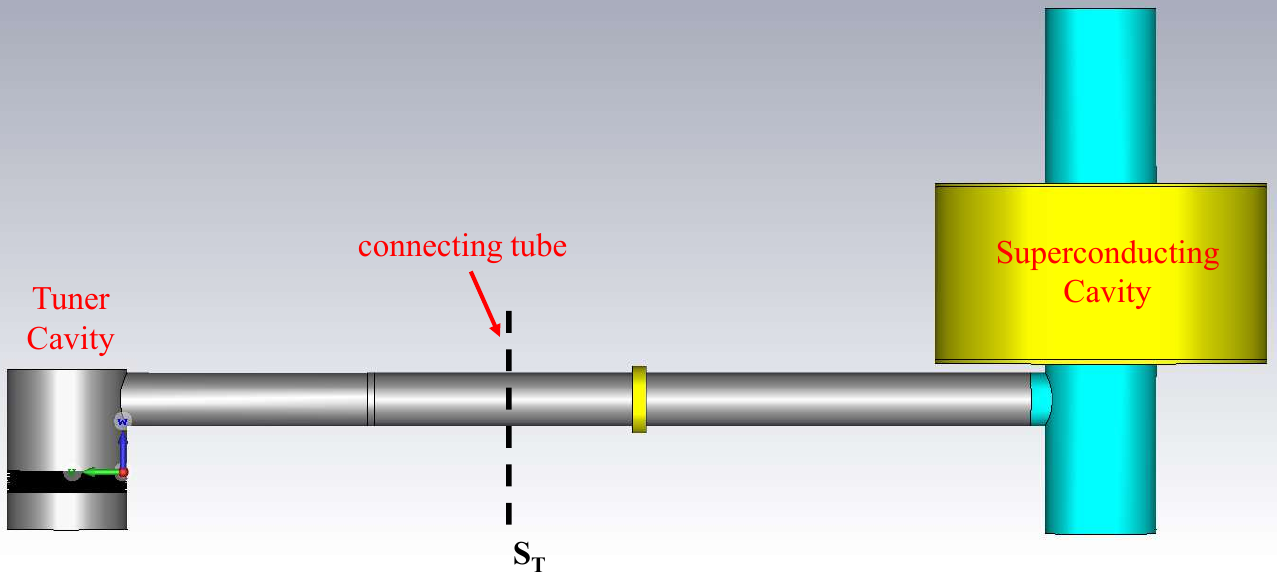}
    \caption{Connection diagram of the superconducting cavity and the tuned cavity}
    \label{fig:crosssection}
\end{figure}

As shown in Fig. \ref{fig:crosssection}, the superconducting cavity and the tuned cavity are connected by a beam tube. The interface of the beam tube is denoted as \(S_{T}\). On both sides of the beam tube, for the \(TM_{010}\) mode of the completely closed cavity, the magnetic field is azimuthal, while the electric field is parallel to the cavity axis. To enable energy transfer between the superconducting and tuned cavities, a perpendicular component of the electric field near the beam tube is necessary. This component, together with the azimuthal magnetic field, generates a Poynting vector along the cavity axis.

Similarly, for the beam tube wall losses, we can consider the additional electric field \(\vec{E}_{\text{rad}}\) at the beam tube \(S_{T}\) as a small perturbation of the intrinsic mode field \(\vec{e}^{(1)}\) of the resonant cavity, where the superconducting cavity is assumed to be completely surrounded by a perfectly conducting surface. The radial field between the superconducting and tuned cavities is the superposition of two fields, each contributed by one of the resonant cavity, and the magnitude of these contributions is proportional to the amplitude of the fundamental mode \(\mathcal{E}^{(1)}\). Thus, the perturbation field at the beam tube can be expressed as:
\begin{equation}
    \left.\vec{E}_{\text{rad}}\right|_{S_{T}} = \mathcal{E}_{S}^{(1)} \left.\vec{e}_{S,T}^{(1)}\right|_{S_{T}} - \mathcal{E}_{T}^{(1)} \left.\vec{e}_{T,S}^{(1)}\right|_{S_{T}}
    \label{eq:radfield}
\end{equation}
where \(\vec{e}_{S, T}^{(1)}\) is the corresponding eigen - mode field of the superconducting cavity at \(S_{T}\), and \(\vec{e}_{T, S}^{(1)}\) is the corresponding eigen - mode field of the tuned cavity at \(S_{T}\).

Therefore, the coupling factor between the superconducting cavity and the tuned cavity is defined as:
\begin{equation}
    \mathcal{K}_{rr} := \frac{c}{\omega_{r}^{(1)}} \int_{S_{T}} \left( \vec{e}_{\text{tube}}^{(1)} \times \vec{h}_{r}^{(1)} \right) \cdot \vec{u}_{z} \, \mathrm{d}s
    \label{eq:couplingfactor}
\end{equation}
where \(\omega_{r}^{(1)}\) is the resonant frequency of the selected mode in the cavity, \(\vec{h}_{r}^{(1)}\) is the corresponding magnetic field mode in the cavity, \(c\) is the speed of light in vacuum, and \(\vec{u}_{z}\) is the unit vector along the cavity axis.

When two cavities are coupled, each $TM_{010}$ mode of an $N$-cell cavity splits into $N$ modes; in a system with $M$ coupled cavities, the mode splits into $M$ eigenmodes. In this work, a single-cell superconducting cavity (fundamental mode only) and a single-mode tuner cavity are employed, leading to the following differential equations for their fundamental modes:

\begin{equation}
    \begin{aligned}
        \frac{d^2 \mathcal{E}_S^{(1)}}{dt^2} 
        &+ \left[ \frac{\omega_0^{(1)}}{Q_0^{(1)}}(1+i) 
        + \frac{\omega_0^{(1)}}{Q_{e,n}^{(1,1)}} \right] 
        \frac{d \mathcal{E}_S^{(1)}}{dt} \\
        &+ \frac{(\omega_0^{(1)})^2}{2} \left( 1 + \mathcal{K}_{r,r} \right) 
        \mathcal{E}_S^{(1)} 
        - \frac{(\omega_0^{(1)})^2}{2} \mathcal{K}_{r,r} 
        \mathcal{E}_{T}^{(1)} \\
        &= -\frac{K_{g}^{(1)}}{\varepsilon_0} \frac{d}{dt} J_{g}
    \end{aligned}
    \label{eq:superconductingcavity}
\end{equation}

\begin{equation}
    \begin{aligned}
        \frac{d^2 \mathcal{E}_T^{(1)}}{dt^2} 
        &+ \left[ \frac{\omega_0^{(1)}}{Q_0^{(1)}}(1+i) 
        + \frac{\omega_0^{(1)}}{Q_{e,n}^{(1,1)}} \right] 
        \frac{d \mathcal{E}_T^{(1)}}{dt} \\
        &+ \frac{(\omega_0^{(1)})^2}{2} \left( 1 + \mathcal{K}_{r,r} \right) 
        \mathcal{E}_T^{(1)} 
        - \frac{(\omega_0^{(1)})^2}{2} \mathcal{K}_{r,r} 
        \mathcal{E}_{S}^{(1)} = 0
    \end{aligned}
    \label{eq:tunercavity}
\end{equation}

\begin{figure}[htbp]
    \centering
    \includegraphics[width=0.45\textwidth]{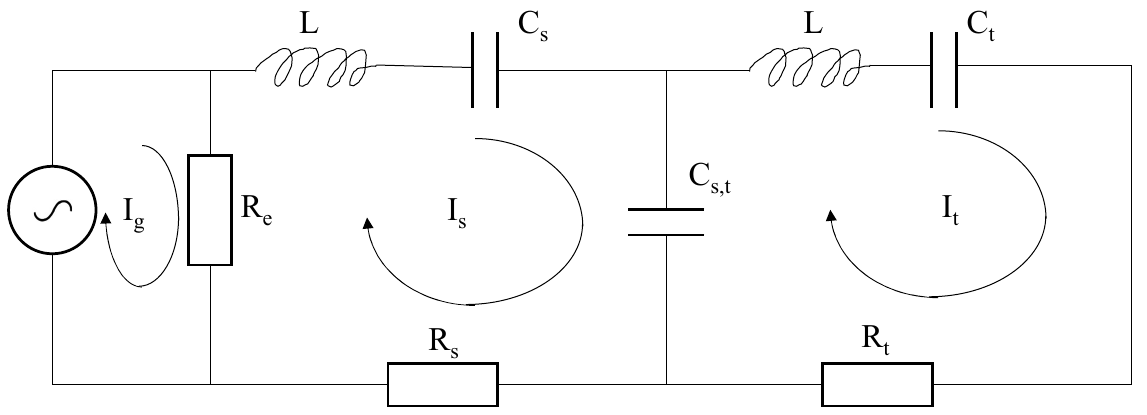}
    \caption{Equivalent circuit model of the cavity tuner system, depicting inductive ($L$), capacitive ($C_s, C_t$), and resistive ($R_s, R_t, R_e$) components.}
    \label{fig:circuitsystem}
\end{figure}

The cavity tuner system is mapped to an equivalent circuit (Fig. \ref{fig:circuitsystem}) through the following definitions:
\begin{equation}
    \begin{cases}
        (\omega_s^{(1)})^2 = \dfrac{1}{LC_s}, \quad (\omega_t^{(1)})^2 = \dfrac{1}{LC_t} \\
        \mathcal{K}_{r,r} = \dfrac{C_s + C_t}{C_{s,t}}, \quad Q_{L,s}^{(1)} = \dfrac{\omega_s^{(1)} L}{R_s + R_e}, \quad Q_{L,t}^{(1)} = \dfrac{\omega_t^{(1)} L}{R_t}
    \end{cases}
    \label{eq:equivalentcircuit}
\end{equation}
where $L$ is the equivalent inductance, $C_s/C_t$ are the capacitances of the superconducting/tuner cavities, $R_s/R_t$ are internal resistances, and $R_e$ is the external load resistance.

Energy equivalence between the field and circuit domains is established by matching the stored energy in a lossless cavity $U_s = \frac{1}{2} \varepsilon_0 |\mathcal{E}_s^{(1)}|^2$ to that in an LC circuit $U_s = \frac{1}{2} L |I_s|^2$, leading to:
\[
\mathcal{E}_s^{(1)} = \sqrt{\frac{L}{\varepsilon_0}} I_s, \quad -\frac{K_g^{(1)}}{\varepsilon_0} J_g = \sqrt{\frac{1}{\varepsilon_0 L}} R_e I_g
\]
Substituting into Kirchhoff's laws yields the circuit equations:
\begin{equation}
    \frac{d^2 I_s}{dt^2} + \frac{R_s}{L} \frac{dI_s}{dt} + \left( \frac{1}{LC_s} + \frac{1}{LC_{s,t}} \right) I_s - \frac{1}{LC_{s,t}} I_t = \frac{R_e}{L} \frac{dI_g}{dt}
    \label{eq:superconductingcircuit}
\end{equation}
\begin{equation}
    \frac{d^2 I_t}{dt^2} + \frac{R_t}{L} \frac{dI_t}{dt} + \left( \frac{1}{LC_t} + \frac{1}{LC_{s,t}} \right) I_t - \frac{1}{LC_{s,t}} I_s = 0
    \label{eq:tunercircuit}
\end{equation}

For the design frequency $\omega_0^{(1)}$, define the nominal capacitance $C = 1/[L(\omega_0^{(1)})^2]$. The superconducting cavity capacitance detunes slightly as:
\[
\frac{C}{C_s} = 1 + \delta_s, \quad \delta_s = 2 \frac{\Delta \omega_s^{(1)}}{\omega_0^{(1)}} + \left( \frac{\Delta \omega_s^{(1)}}{\omega_0^{(1)}} \right)^2, \quad \Delta \omega_s^{(1)} = \omega_s^{(1)} - \omega_0^{(1)}
\]
The circuit coupling coefficient is:
\[
k_{r,r} = \frac{C}{C_{s,t}} = \frac{(\omega_s^{(1)})^2}{(\omega_0^{(1)})^2} \frac{\mathcal{K}_{r,r}}{2}
\]

In matrix form, the system equation becomes:
\begin{equation}
    \frac{1}{(\omega_0^{(1)})^2} \ddot{\vec{I}} + \mathbf{B} \dot{\vec{I}} + \mathbf{A} \vec{I} = \frac{\omega_1^{(1)}}{(\omega_0^{(1)})^2 Q_{e,1}^{(1,1)}} \dot{\vec{I}}_g
    \label{eq:matrixform}
\end{equation}
with:
\[
\vec{I} = \begin{bmatrix} I_s \\ I_t \end{bmatrix}, \quad \vec{I}_g = \begin{bmatrix} I_g \\ 0 \end{bmatrix},
\]

\[
\mathbf{A} = \begin{bmatrix} 1 + \delta_s + k_{r,r} & -k_{r,r} \\ -k_{r,r} & 1 + \delta_t + k_{r,r} \end{bmatrix}, 
\label{eq:matrixA}
\]

\[
\mathbf{B} = \frac{1}{(\omega_0^{(1)})^2} \begin{bmatrix} \omega_s^{(1)}/Q_s^{(1)} & 0 \\ 0 & \omega_t^{(1)}/Q_t^{(1)} \end{bmatrix}
\]

For RF-driven steady states, separate the fast oscillating term $e^{i\omega_g t}$:
\[
J_g(t) = \hat{J}_g e^{i\omega_g t}, \quad \mathcal{E}^{(1)}(t) = \hat{\mathcal{E}} e^{i\omega_g t}, \quad I(t) = \hat{I} e^{i\omega_g t}
\]
Assuming slow-varying envelopes ($\dot{\hat{I}} \ll \omega_g \hat{I}$), the first-order approximation yields:
\[
\dot{\vec{I}} + \frac{(\omega_0^{(1)})^2}{2} \mathbf{B} \vec{I} + \frac{i\omega_g}{2} \left[ 1 - \left( \frac{\omega_0^{(1)}}{\omega_g} \right)^2 \mathbf{A} \right] \vec{I} = - \frac{i\omega_1^{(1)}}{2\omega_g Q_{e,1}^{(1,1)}} \vec{I}_g e^{-i\omega_g t}
\]
For constant-amplitude steady states ($\vec{\hat{J}}_g = \text{const}$), the solution satisfies:
\[
\frac{i\omega_g}{2} \left[ 1 - \left( \frac{\omega_0^{(1)}}{\omega_g} \right)^2 \mathbf{A} - \frac{i(\omega_0^{(1)})^2}{\omega_g} \mathbf{B} \right] \vec{\hat{\mathcal{E}}} = - \frac{K_g^{(1)}}{2\varepsilon_0} \vec{\hat{J}}_g
\]

The eigenvalue problem for the cavity tuner system is expressed as:
\begin{equation}
    \mathbf{A} \vec{v}^{(j)} = \Omega^{(j)} \vec{v}^{(j)}
    \label{eq:eigenvalue}
\end{equation}
where $\mathbf{A}$ is the system matrix defined in Eq. \eqref{eq:matrixA}, $\vec{v}^{(j)}$ are the eigenvectors, and $\Omega^{(j)}$ are the eigenvalues. The relationship between the eigenvalue and the system's eigenfrequency is:
\begin{equation}
    \Omega^{(j)} := \frac{(\omega^{(j)})^2}{(\omega_0^{(1)})^2}
    \label{eq:eigenfrequency}
\end{equation}
Here, $\omega^{(j)}$ represents the $j$-th order eigenfrequency of the coupled system, and $\omega_0^{(1)}$ is the design frequency of the superconducting cavity's fundamental mode.

\begin{figure}[htbp]
    \centering
    \includegraphics[width=0.45\textwidth]{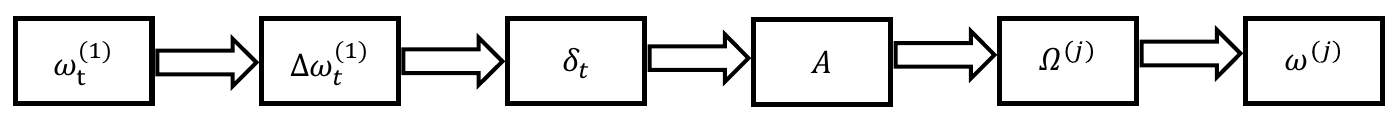}
    \caption{Tuning principle of the cavity tuner system: mechanical/electromagnetic adjustment of the tuner cavity frequency modifies the coupled system's eigenfrequencies to match the RF source.}
    \label{fig:tuningprinciple}
\end{figure}

The tuning principle of the cavity tuner is illustrated in Fig. \ref{fig:tuningprinciple}: by adjusting the resonant frequency $\omega_t^{(1)}$ of the tuner cavity, the frequency detuning $\Delta \omega_t^{(1)} = \omega_t^{(1)} - \omega_0^{(1)}$ is modified, which in turn alters the detuning parameter $\delta_t$ (defined analogous to $\delta_s$ for the superconducting cavity). This change propagates to the system matrix $\mathbf{A}$ (see Eq. \eqref{eq:matrixA}), shifting the overall eigenfrequencies of the coupled system to match the RF power source frequency.

\begin{figure}[htbp]
    \centering
    \includegraphics[width=0.45\textwidth]{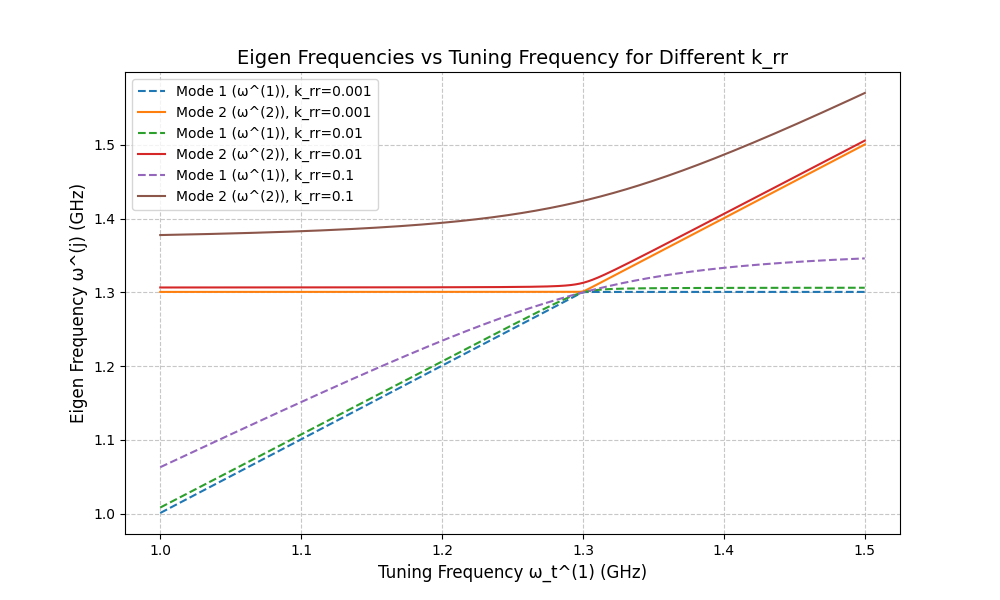}
    \caption{System eigenfrequency shifts as a function of tuner cavity frequency for varying coupling coefficients, showing broadened tuning range with increased $k_{r,r}$.}
    \label{fig:eigenfrequencies}
\end{figure}

As a numerical example, consider a superconducting cavity with a fixed resonant frequency and design frequency of $1.3$ GHz, paired with a tuner cavity adjustable from $1.0$ to $1.5$ GHz via a mechanical actuator. Fig. \ref{fig:eigenfrequencies} depicts the system's eigenfrequency responses for coupling coefficients $k_{r,r} = 0.001$, $0.01$, and $0.1$. The curves show that higher coupling coefficients broaden the tuning range while maintaining frequency selectivity, demonstrating the trade-off between tuning agility and mode purity.

\section{Advances and Potential of Cavity Tuners}

Currently, the application of cavity tuners remains in the research phase, with a primary emphasis on simulation design and theoretical analysis. According to existing literature, notable application examples of cavity tuners are mainly derived from Fermilab's simulation-based design of the Quarter-Wave Resonator (QWR) for the PIP-III project \cite{Kuzikov2022}.

\begin{figure}[htbp]
    \centering
    \includegraphics[width=0.45\textwidth]{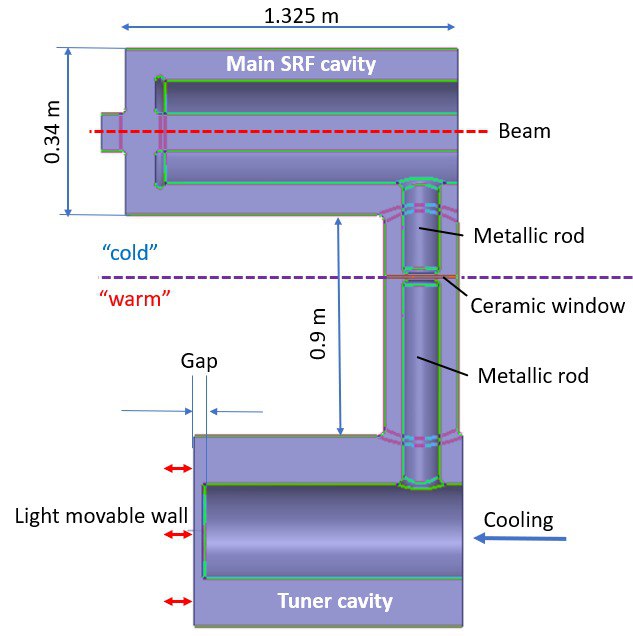}
    \caption{QWR cavity tuner design in the PIP-III project}
    \label{fig:QWR}
\end{figure}

In the PIP-III project, Fermilab put forward a QWR design scheme centered around cavity tuners. This innovative approach achieves precise regulation of the superconducting cavity's frequency by externally coupling a tuner cavity to the superconducting cavity. Simulation results have revealed that cavity tuners can significantly enhance the tuning speed and precision of QWRs while ensuring a wide tuning range. This design offers novel concepts and methodologies for tuning superconducting cavities in future high-energy particle accelerators.

Despite the current limitations in the practical application of cavity tuners, their successful cases in simulation design showcase considerable potential, particularly for emerging superconducting materials. This tuning approach is especially advantageous for future Nb$_{3}$Sn-based superconducting cavity, which exhibit poor mechanical ductility. Traditional mechanical tuning methods, relying on physical deformation of the cavity, may damage the delicate surface of Nb$_{3}$Sn coatings, compromising their superconducting properties. In contrast, the non-contact nature of cavity tuners avoids imposing mechanical stress on the superconducting cavity while achieving precise frequency matching with the power source. With continuous technological development and refinement, cavity tuners are anticipated to be adopted in a wider array of superconducting accelerator projects, thereby further boosting the performance and stability of superconducting cavity.

Cavity tuners present several distinct advantages over traditional tuning systems:
\begin{itemize}
    \item \textbf{High-speed Tuning}: Cavity tuners accomplish frequency adjustments by modifying the frequency of an external resonant cavity. This mechanism enables them to achieve much faster tuning speeds compared to traditional mechanical tuners, facilitating rapid responses to frequency-related requirements.
    \item \textbf{Precise Tuning}: Leveraging electromagnetic tuning techniques, cavity tuners can offer extremely high tuning accuracy. This allows for the meticulous control of the superconducting cavity's frequency, which is crucial for optimizing the cavity's performance.
    \item \textbf{Broad Tuning Spectrum}: Cavity tuners feature a large tuning range that spans a wide frequency spectrum. This characteristic enables them to meet diverse operational needs, providing greater flexibility in different accelerator applications.
    \item \textbf{Streamlined Structure}: The structure of cavity tuners is relatively straightforward, which simplifies the manufacturing and maintenance processes. This not only reduces the overall complexity of the system but also cuts down on associated costs.
    \item \textbf{Enhanced Reliability}: Since cavity tuners employ a non - contact tuning method, they minimize the wear and tear on mechanical components. As a result, the system's reliability is improved, and its service life is extended.
\end{itemize}

In summary, cavity tuners excel in terms of tuning speed, accuracy, range, structural simplicity, and reliability. These outstanding features suggest a highly promising future for their application in the field of superconducting accelerators.

\section{Design of cavity tuner}

The cavity tuner constitutes a dual-component system comprising a superconducting cavity and a tuner cavity. Its design necessitates a holistic consideration of multiple interdependent factors, including cavity geometry, coupling topology, tuning actuation, and thermal management, which are systematically detailed in the following subsections.  

\subsection{Cavity Design}  

The architectural design of the tuner system forms the basis for its operational performance, encompassing the structural configurations of both the superconducting and tuner cavities. The superconducting cavity employed in this study adheres to the widely adopted 1.3~GHz single-cell design paradigm \cite{Meena2022ebx}, with detailed geometric parameters and surface treatment protocols omitted here as they are beyond the scope of this paper. The tuner cavity is engineered as a room-temperature coaxial resonator positioned external to the cryogenic system, leveraging its cylindrical symmetry to facilitate efficient frequency tuning and thermal management.  

\subsubsection{Tuner Cavity Architecture}  
\begin{figure}[htbp]
    \centering
    \includegraphics[width=0.45\textwidth]{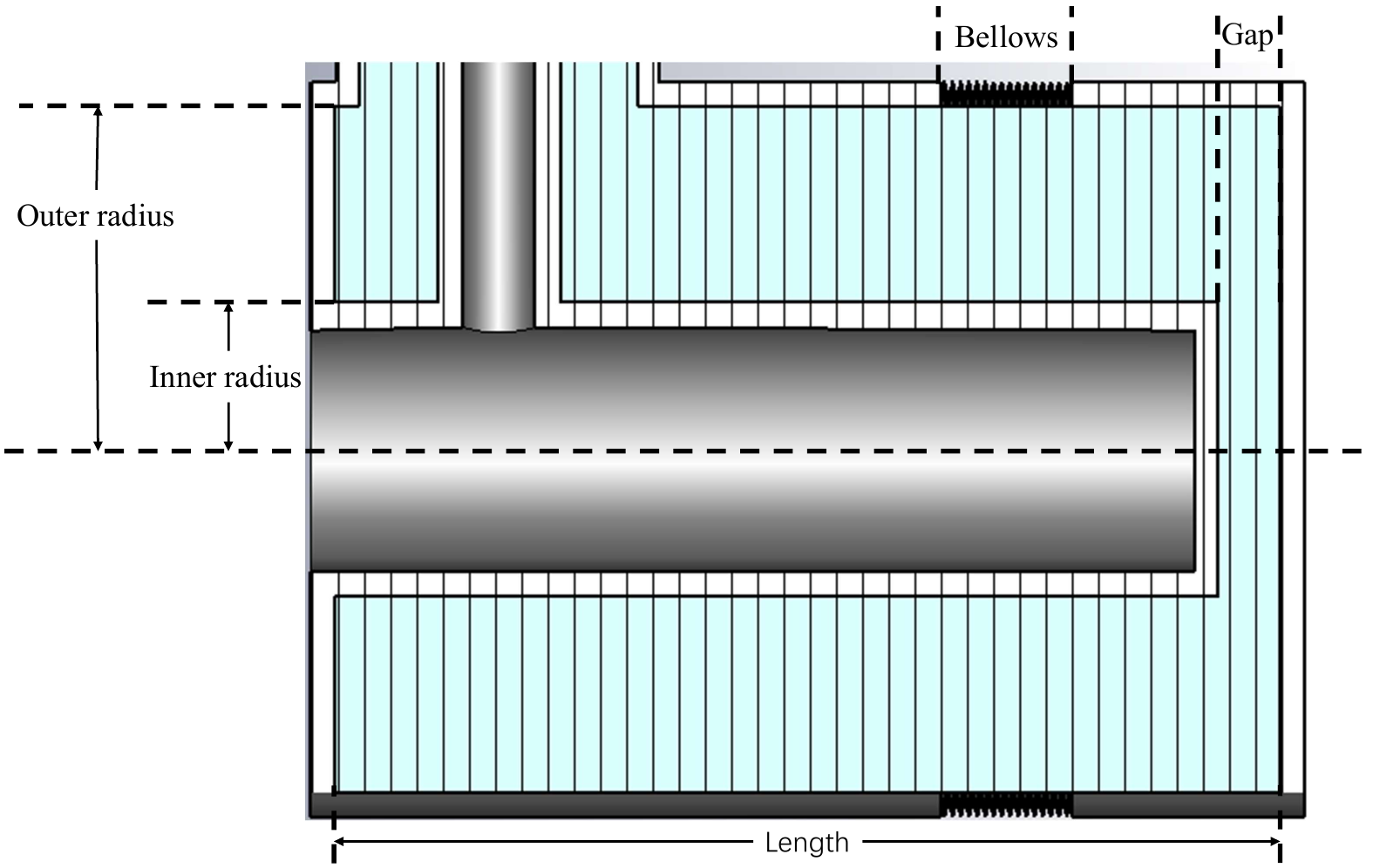}
    \caption{Cross-sectional view of the tuner cavity, depicting its coaxial architecture and functional components for frequency tuning and thermal management.}
    \label{fig:tunercrosssection}
\end{figure}

The cross-sectional geometry of the tuner cavity is illustrated in Fig. \ref{fig:tunercrosssection}, featuring a segmented coaxial structure optimized for microwave coupling and mechanical adjustability:  
The upper section establishes a rigid coaxial connection with the superconducting cavity, ensuring low-loss transmission of electromagnetic power. The inner conductor extends into the pipe to form a probe coupler, enabling efficient energy transfer between the two cavities.  
A bellows-integrated outer conductor introduces mechanical flexibility, allowing precise longitudinal displacement ($\pm 5 \sim mm$) of the cavity endwall. Finite element analysis indicates a frequency sensitivity of $2.4 \sim MHz/mm$ at the 1.3GHz design frequency, providing fine-grained control over the tuner cavity's resonant frequency.  
The mechanical actuation system, located at the right port, employs a electromagnetic-driven linear stage with sub-micron positional accuracy ($ \pm 0.5\sim \mu m$) to modulate the cavity length via the bellows mechanism. This enables both coarse frequency adjustments and dynamic compensation for thermal drift in the superconducting cavity.  
The left port houses a cooling interface for a dual-loop system (air or liquid), designed to maintain the tuner cavity wall temperature at $ \sim 25 \pm 1  ^{\circ}C$. This mitigates thermal expansion effects, which could otherwise induce frequency shifts with an estimated thermal coefficient of less than 50~Hz/°C.  
A dedicated vacuum extraction port at the bottom ensures an interior pressure below $ 10^{-6} \sim $mbar during operation, minimizing dielectric losses and preserving the cavity's high-quality factor ($Q_{t} > 2 \times 10^{4} $ at room temperature).  

\subsubsection{Electromagnetic Field Distribution}  
\begin{figure}[htbp]
    \centering
    \subfigure[Electric Field Distribution]{%
        \includegraphics[width=0.45\textwidth]{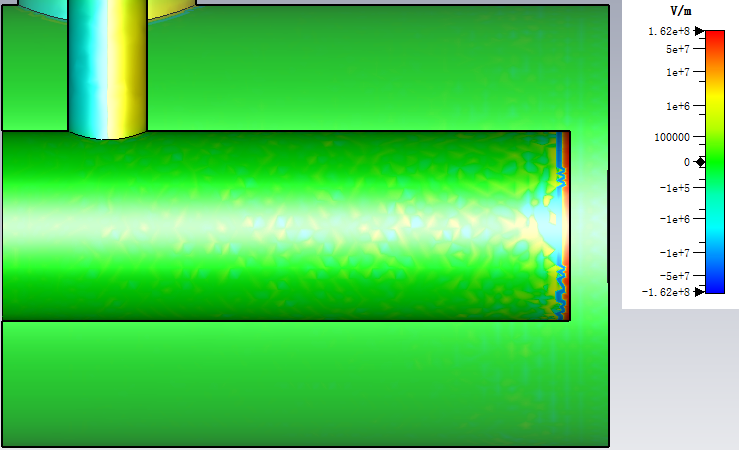}%
        \label{fig:electricalfield}%
    }
    \subfigure[Magnetic Field Distribution]{%
        \includegraphics[width=0.45\textwidth]{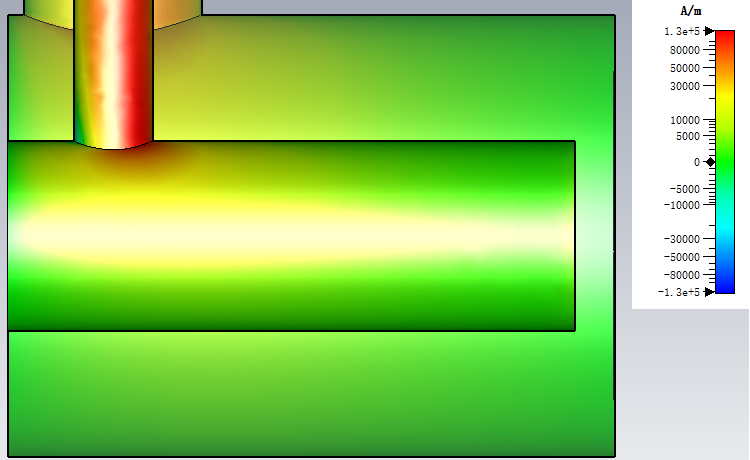}%
        \label{fig:magneticfield}%
    }
    \caption{Electromagnetic field distributions within the tuner cavity, showing E-field concentration at the coaxial gap and H-field dominance in the annular region.}
    \label{fig:fielddistribution}
\end{figure}

Computational simulations of the tuner cavity reveal distinct electric and magnetic field distributions critical to its operational efficiency, as shown in Fig. \ref{fig:fielddistribution}:  
The electric field (E-field) is concentrated at the coaxial gap between the inner conductor and the adjustable endwall, forming a high-field region. This localization enhances the sensitivity of frequency tuning to mechanical displacement, enabling precise adjustments.  
The magnetic field (H-field) predominates in the annular space between the inner conductor and outer wall, where it interacts minimally with the cooling channels to reduce eddy current losses. The symmetric distribution of the H-field facilitates uniform thermal dissipation through the outer conductor, ensuring stable operation under varying power loads.  

This coaxial design optimizes the trade-off between tuning resolution, thermal stability, and microwave performance. The cylindrical symmetry suppresses higher-order mode excitation, while the modular architecture allows independent optimization of mechanical, thermal, and electromagnetic subsystems, ensuring robust operation across the intended frequency range.  
\subsection{Coupling Design}

The superconducting cavity and the tuner cavity are interconnected via a feedline. At the superconducting cavity end, the inner conductor of the feedline functions as a probe, which is inserted into the beam - pipe of the superconducting cavity. This configuration allows for efficient coupling with the interior electromagnetic field of the superconducting cavity. At the tuner cavity end, the inner and outer conductors of the feedline are respectively connected to the corresponding inner and outer conductors of the tuner cavity. Through the interaction of these conductors, energy coupling between the superconducting cavity and the tuner cavity is achieved.

Given that the superconducting cavity is situated inside the cryostat and the tuner cavity is outside, a coupling window is essential for effective coupling. The design of this coupling window must take into account multiple factors such as coupling efficiency, power loss, and thermal conduction. These considerations are crucial to guarantee seamless energy transfer between the two cavities.

\subsubsection{Coupling Window Design}

\begin{figure}[htbp]
    \centering
    \includegraphics[width=0.45\textwidth]{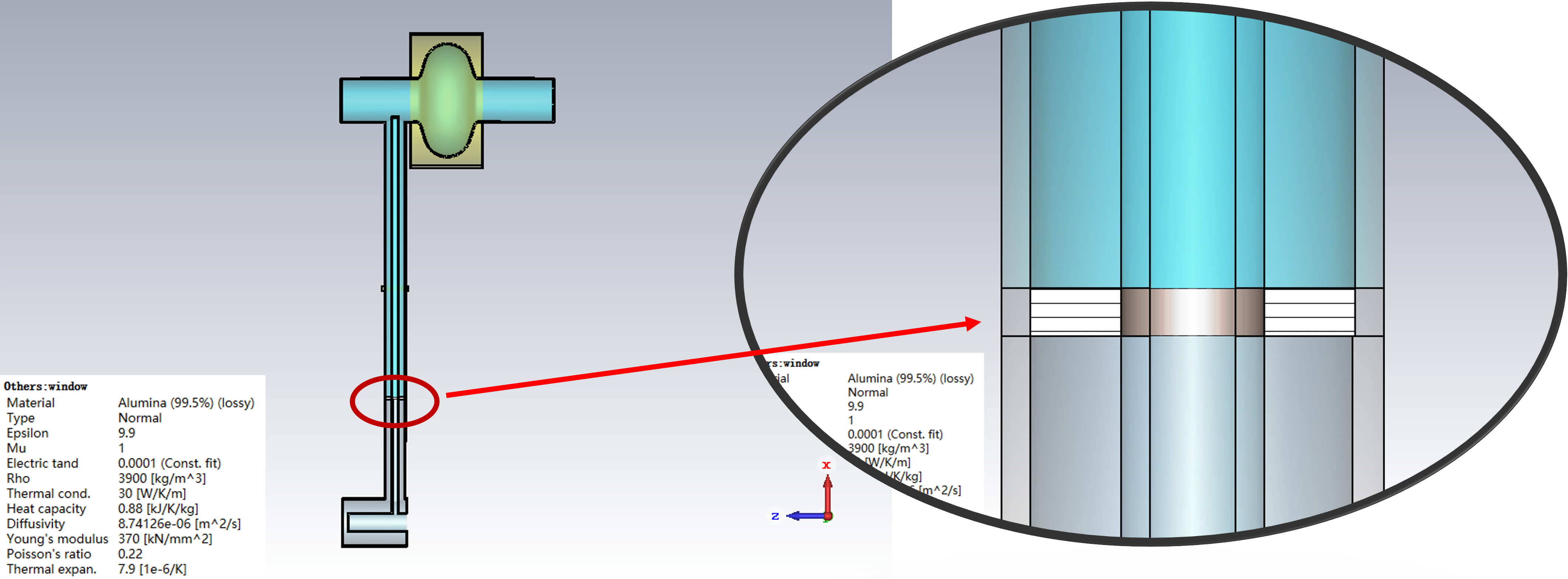}
    \caption{Coupling Window Design Diagram}
    \label{fig:windowdesign}
\end{figure}

The coupling window serves a dual purpose. Firstly, it isolates the vacuum environment inside and outside the cryostat for the tuner cavity. Secondly, it minimizes microwave reflection and loss to ensure effective coupling between the superconducting and tuner cavities. To reduce loss, alumina ceramic is selected as the material for the window. As depicted in Figure \ref{fig:windowdesign}, the window has a ring - shaped structure. The inner ring is connected to the inner conductor of the feedline, and the outer ring is connected to the outer conductor. This setup effectively isolates the vacuum regions while allowing for microwave transmission.

\begin{figure}[htbp]
    \centering
    \includegraphics[width=0.45\textwidth]{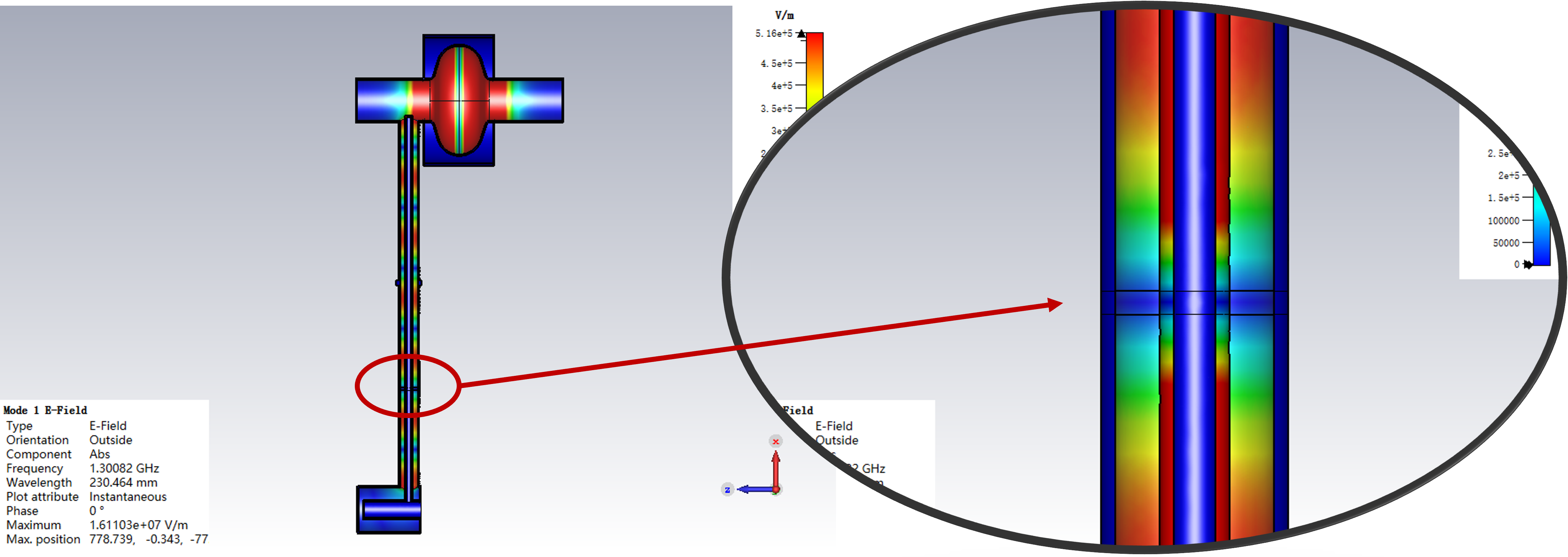}
    \caption{Coupling Window Position Design Diagram}
    \label{fig:windowposition}
\end{figure}

The inner conductor of the window is susceptible to overheating, which can lead to stress, deformation, and in severe cases, voltage breakdown and secondary electron multiplication due to excessive local electric fields. To mitigate these issues, the window should be positioned at the antinode of the wave, as shown in Figure \ref{fig:windowposition}.

\subsubsection{Coupling Factor Calculation}

The coupling factor is a key parameter in the cavity tuner system, quantifying the energy transfer efficiency between the superconducting cavity and the tuner cavity. Its value directly impacts the efficiency and stability of the tuning system.

\begin{figure}[htbp]
    \centering
    \includegraphics[width=0.45\textwidth]{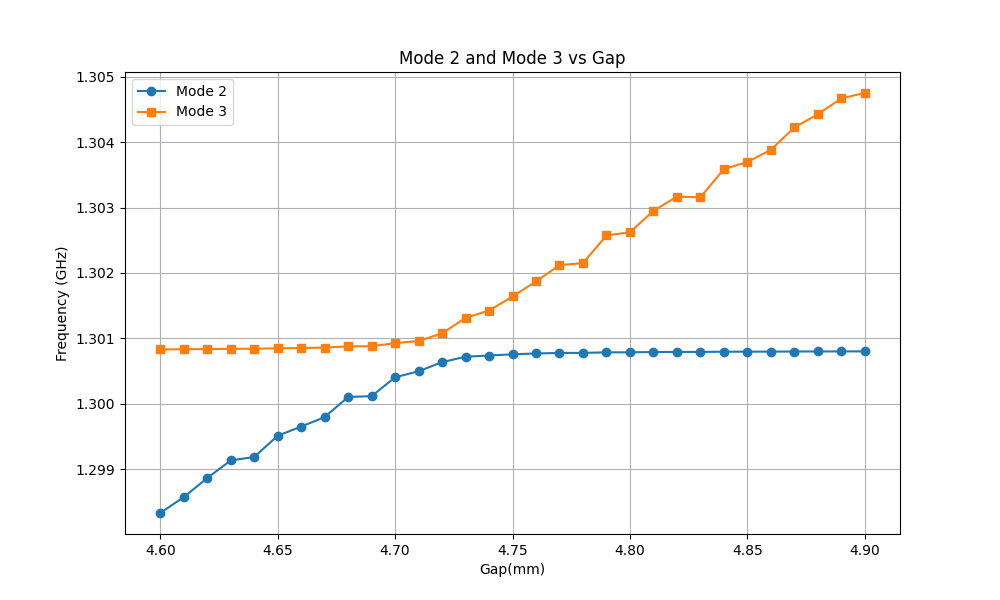}
    \caption{Coupling factor curve calculated by CST simulation}
    \label{fig:couplingfactor}
\end{figure}

By utilizing the CST software to simulate the electric field distribution of the superconducting and tuner cavities, the eigenmode field distributions of both cavities can be obtained. By comparing these simulation results with theoretical values, the coupling factor can be accurately calculated. Figure \ref{fig:couplingfactor} presents the variation curves of two eigenfrequencies when the tuner cavity gap varies from 4.6 to 4.9.

The magnitude of the coupling factor is influenced by the position of the coupling port of the superconducting cavity and the insertion depth of the coupling probe. By carefully adjusting these two parameters, the coupling factor can be effectively tuned, thereby achieving high-efficiency coupling between the two cavities.

From Figure \ref{fig:couplingfactor}, it can be inferred that the closer the coupling port of the superconducting cavity is to the iris, the larger the coupling factor, indicating a higher energy transfer efficiency between the superconducting cavity and the tuner cavity.

\subsection{Mechanical Tuning Design of the Tuner Cavity}
\label{sec:mechanicaltuning}

The mechanical tuning design of the tuner cavity employs an electromagnetic coil - driven approach \cite{kashikhin_electromagnetic_2009}. This system mainly consists of two coils: the main coil and the moving coil. The main coil is responsible for generating a static radial magnetic field. When the moving coil is placed in this magnetic field, it experiences a force that causes longitudinal displacement. This displacement changes the length of the cavity, thereby adjusting its resonant frequency.

\begin{figure}[htbp]
    \centering
    \includegraphics[width=0.45\textwidth]{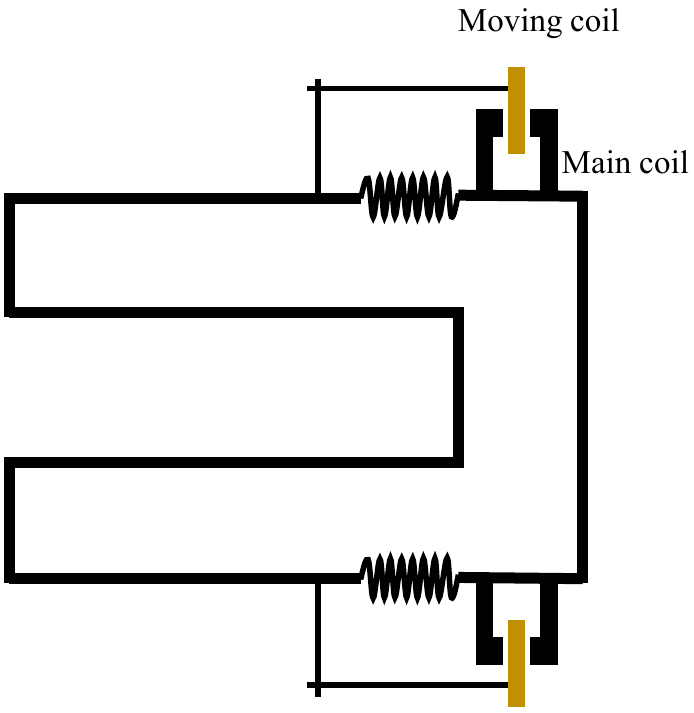}
    \caption{Mechanical tuning design of the tuner cavity}
    \label{fig:mechanicaltuning}
\end{figure}

\begin{figure}
    \centering
    \includegraphics[width=0.45\textwidth]{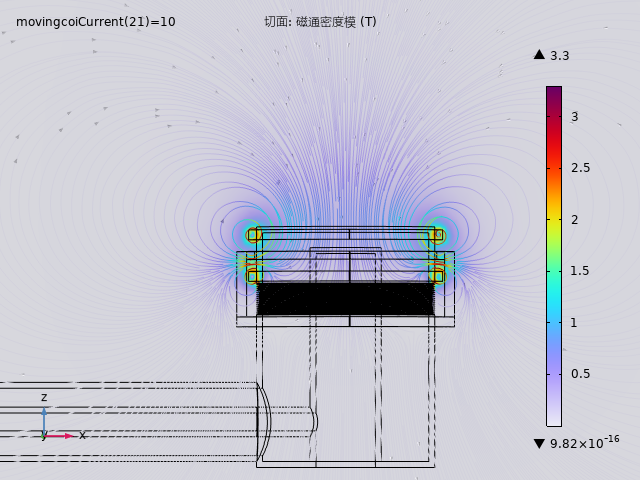}
    \caption{Magnetic field distribution of the mechanical tuning system for the tuner cavity}
    \label{fig:magneticfieldlines}
\end{figure}

As shown in Figure \ref{fig:mechanicaltuning}, the main coil is connected to a DC power supply. When energized, it generates a static magnetic field, as shown in Figure \ref{fig:magneticfieldlines}. The moving coil, under the influence of the force \(F = I\cdot L\cdot B\) generated by the current - induced magnetic field, undergoes longitudinal displacement, which in turn changes the length of the tuner cavity. By precisely adjusting the magnitude and direction of the current in the main and moving coils, accurate tuning of the tuner cavity can be achieved.

\begin{figure}[htbp]
    \centering
    \includegraphics[width=0.45\textwidth]{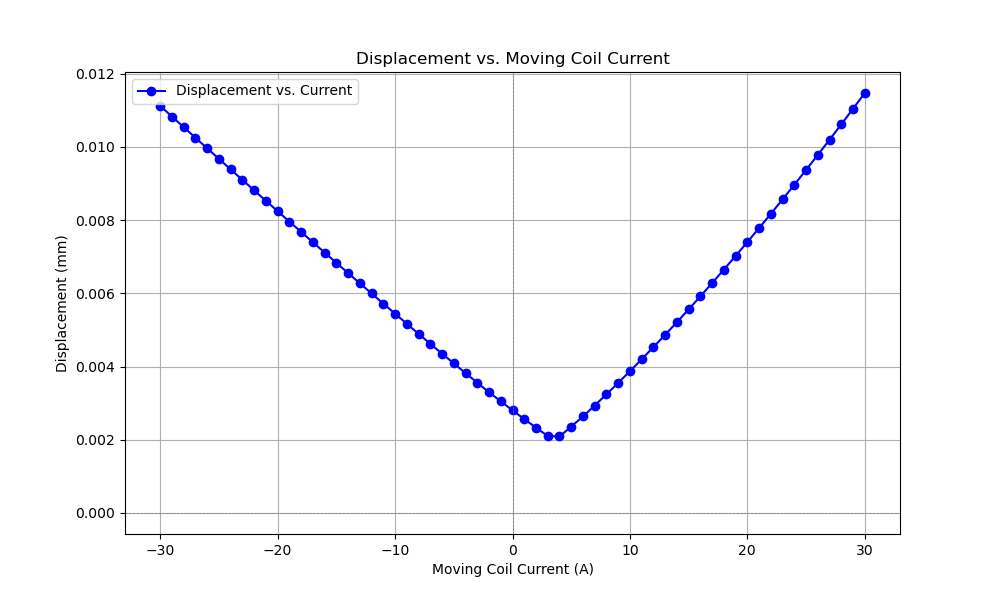}
    \caption{Mechanical tuning characteristic curve of the tuner cavity}
    \label{fig:mechanicaltuningcurve}
\end{figure}

Using the magnetic field and solid mechanics modules in COMSOL, a simulation of the mechanical tuning system of the tuner cavity was conducted. The resulting mechanical tuning characteristic curve of the tuner cavity is presented in Figure \ref{fig:mechanicaltuningcurve}.

\subsection{Cooling Design of the Cavity}

During operation, both the superconducting cavity and the tuner cavity generate heat. To ensure their proper functioning and maintain optimal performance, an effective cooling mechanism is required. The superconducting cavity is cooled using liquid helium due to its extremely low-temperature requirements for superconducting operation. The tuner cavity, on the other hand, can be cooled using either gas or water cooling methods, which offer flexibility based on the specific operating conditions and system requirements.

\section{Conclusion}
\label{sec:conclusion} 

This study introduces a novel cavity tuner for 1.3GHz single-cell superconducting radio-frequency (SRF) cavity, addressing the trade-offs between tuning speed, precision, and range in conventional systems. By employing an external coaxial tuner cavity coupled to the SRF cavity via a probe-based architecture, the design enables efficient energy transfer and mechanical adjustability, supported by theoretical analysis and computational validation.  

The core innovations include:  
\begin{itemize}
    \item A coaxial feedline integrated with a ring-shaped alumina ceramic coupling window, which minimizes microwave reflection and thermal conduction while isolating the cryogenic vacuum environment. Positioning the window at the electric field antinode reduces the risk of voltage breakdown by suppressing local field enhancement, a critical improvement for high-power operation.  
    \item An electromagnetic coil-driven mechanical tuning system that leverages the Lorentz force (\(F = I \cdot L \cdot B\)) to achieve a tuning speed of $ \sim MHz/s $ and sub-micron positional accuracy. This mechanism provides a frequency sensitivity of $\sim 2.4 kHz/mm$, enabling a wide tuning range of $ \sim \pm 15 kHz $—three times broader than traditional mechanical tuners. 
    \item A thermally optimized coaxial structure that concentrates electric fields at the coupling gap for efficient mode excitation and symmetrically distributes magnetic fields to minimize eddy current losses (room-temperature quality factor \(Q_t > 2 \times 10^4\)). A dual-loop cooling system maintains the tuner cavity at $ \sim 25 ^{\circ} C $, ensuring thermal drift and stable operation under dynamic loads. 
\end{itemize}

CST simulations validate the coupling mechanism, showing that adjusting the probe depth and port position can optimize the coupling factor \(\mathcal{K}_{rr}\). COMSOL multi-physics modeling confirms the mechanical system’s robustness, surpassing the precision of existing piezoelectric and stepper-motor tuners. The design's non-contact operation offers significant advantages for brittle Nb$_{3}$Sn-based SRF cavity, eliminating mechanical stress on surface coatings and preserving superconducting performance. 

Future research will focus on adapting the design for 1.3GHz 9-cell high-frequency cavity and integrating model-based control algorithms to address dynamic detuning in pulsed modes. The modular architecture also paves the way for multi-cell cavity systems, positioning this tuner as a key enabling technology for next-generation light sources and colliders, such as the Energy Recovery Linac and Circular Electron-Positron Collider.

\bibliographystyle{elsarticle-num-names} 
\bibliography{bibfileTemplate}

\end{document}